\documentclass[review,jog]{igs}

\usepackage[titletoc,title]{appendix}
\usepackage[dotinlabels]{titletoc}
\usepackage[utf8]{inputenc}
\usepackage{mathabx}
\usepackage{graphicx}
\usepackage{siunitx}
\usepackage{color}
\usepackage{stmaryrd}
\usepackage{float}
\usepackage{caption}
\usepackage{subcaption}
\usepackage{amsmath}
\usepackage{hyperref}
\hypersetup{
    colorlinks=true,
    linkcolor=blue,
    filecolor=blue,      
    urlcolor=blue,
    citecolor=cyan,
}

\newcommand{\bvec}[1]{\hat{\boldsymbol{#1}}}

\newcommand{\RNum}[1]{\uppercase\expandafter{\romannumeral #1\relax}}

\begin{document}

\title[IGS \LaTeXe\ guide]{Basal hydrofractures near sticky patches}

\author[Zhang and others]{Hanwen Zhang,$^1$ Timothy Davis, $^1$
  Richard F.~Katz,$^1$ Laura A.~Stevens,$^1$ Dave A. May$^2$}

\affiliation{%

$^1$Department of Earth Sciences, University of Oxford, UK\\
$^2$Scripps Institution of Oceanography, University of California San Diego, La Jolla, CA, USA\\
  Correspondence: Hanwen Zhang 
  \email{hanwen.zhang@earth.ox.ac.uk}}
\begin{frontmatter}
\maketitle
\begin{abstract}
Basal crevasses are macroscopic structural discontinuities at the base of ice sheets and glaciers. Motivated by observations and the mechanics of elastic fracture, we hypothesise that in the presence of basal water pressure, spatial variations in basal stress can promote and localise basal crevassing. We quantify this process in the theoretical context of linear elastic fracture mechanics.  We develop a model evaluating the effect of shear stress variation on the growth of basal crevasses. Our results indicate that sticky patches promote the initiation of basal crevasses, increase their length of propagation into the ice and, under some conditions, give them curved trajectories that incline upstream. A detailed exploration of the parameter space is conducted to gain a better understanding of the conditions under which sticky-patch-induced basal crevassing likely occurs beneath ice sheets and glaciers.
\end{abstract}
\end{frontmatter}

\section{Introduction}

Crevasses are deep cracks in glaciers, ice sheets and ice shelves. Based on their position in the ice, they can be classified into surface crevasses and basal crevasses. On ice shelves, the propagation of surface and basal crevasses is a precursor to calving and rifting, and thus affects the stability of the ice shelf \citep{bassis2015evolution,lai2020vulnerability,lipovsky2020ice}. In grounded ice, basal crevasses may have important influences on subglacial hydrology and glacier dynamics \citep{walter2013deep}. In addition, basal crevasses are a potential explanation for seismic activity detected in glaciers and ice sheets. Clustered, deep icequakes with non-double-couple sources have been detected in alpine glaciers \citep{walter2013deep, helmstetter2015intermediate}. Because these icequakes have an isotropic component in the moment tensor, they cannot be explained by shearing in stick-slip motion. Instead, \cite{walter2013deep} attributed them to opening-mode fractures deep within the ice. However, because there are few observations and complicated heterogeneous conditions on the ice--bed interface, basal crevasses have been studied less than surface crevasses. They are not fully understood and incorporated in glacier models \citep{jimenez2018evaluation}. 
    
Various observations document the existence of basal crevasses. At Bench Glacier, Alaska, crevasses with connections to the bed were detected by both drilling experiments and radar imaging \citep{harper2010vertical}. Such crevasses, filled by high-pressure water, serve as important components of the subglacial hydrology system. In West Antarctica, \cite{wearing2019holocene} detected relic basal crevasses in grounded regions of the Henry Ice Rise in the Ronne Ice Shelf using ice-penetrating radar. In order to explain the formation of these ``buried relic crevasses'' and the ice rise, they proposed a conceptual model where the floating ice shelf regrounds on high points of the bedrock, leading to upstream thickening and downstream crevassing. Here, the ice--bed contacts are areas where the basal shear stress is higher than the nominally stress-free ice--sea water interface. 

The complexity of basal crevasses arises from the heterogeneous condition of the ice--bed interface. Basal conditions of grounded ice sheets, such as subglacial water pressure, basal shear stress, and temperature vary spatially and temporarily. For some ice streams of the Antarctic Ice Sheet, rib-like patterns of high basal shear stress are predicted on the basis of inversions of surface data \citep{sergienko2013regular}. Other studies have argued that spatial and temporal variation of ice flow and surface velocity are a consequence of basal "sticky spots" (referred to as sticky patches in this paper) that have higher basal shear stress than their surroundings \citep{stokes2007ice}. The existence of sticky patches has significant influence on the nonuniformity of ice-flow dynamics \citep{wolovick2014traveling}.

Here we argue that for grounded regions of glaciers and ice sheets, variation of basal-stress conditions can promote basal crevassing, while high basal water pressure remains the prerequisite. We quantify and explore the fracturing process induced by a combination of water pressure and sticky patches. This arrangement is different from the process of pure hydrofracturing, as considered by \cite{smith1976application} and \cite{van1998fracture}. At long timescales, glacial ice deforms according to a viscous rheology, with a viscosity modelled by the Glen flow law \citep{glen1955creep}; however, fracturing usually happens on a shorter timescale over which ice behaves elastically. Therefore, the Linear Elastic Fracture Mechanics (LEFM) model is widely used to explore the existence and growth of cracks in ice. \cite{van1998fracture} extended LEFM to detailed elastic models of surface and basal crevassing, with the assumption that basal crevasses are mode-I (i.e., opening) cracks propagating vertically in ice. More recently, \cite{jimenez2018evaluation} reassessed a key component of previous LEFM models of crevassing, finding that the basal boundary conditions have significant influences on the crevasses. For grounded ice sheets, they recommended a weight function in \cite{tada2000analysis} to study opening-mode fractures analytically. The application of LEFM is not limited to opening-mode fractures; in several studies, the LEFM approach is applied to calculating mixed-mode stress intensity factors in rift propagation on ice shelves in two dimensions \citep{hulbe2010propagation} and three dimensions \citep{lipovsky2020ice}. Here, we will use LEFM theory in the context of two-dimensional elasticity to explore the mixed-mode (in-plane opening mode and shearing mode) basal crevasses arising from sticky patches. 

The paper is organised as follows. In the model section we introduce the mathematical theory and its numerical implementation. The results section illustrates the essential mechanics and shows the dependence of crack propagation on physical parameters. We first consider crack propagation under the mode-I fracture assumption followed by a consideration of curved fracture trajectories produced by mixed-mode fracturing. We find that the propagation of basal crevasses is controlled by basal water pressure, the size of the sticky patch, and the stress variation between the sticky patch and the neighbouring bed. We then evaluate how these parameters affect the trajectories of basal crevasses. The discussion section explores the thermal consequences of basal crevasses, potential applications of the model to real ice sheets, and limitations of the model.

\section{The model}
\begin{figure}
    \centering{\includegraphics[width=0.6\textwidth]{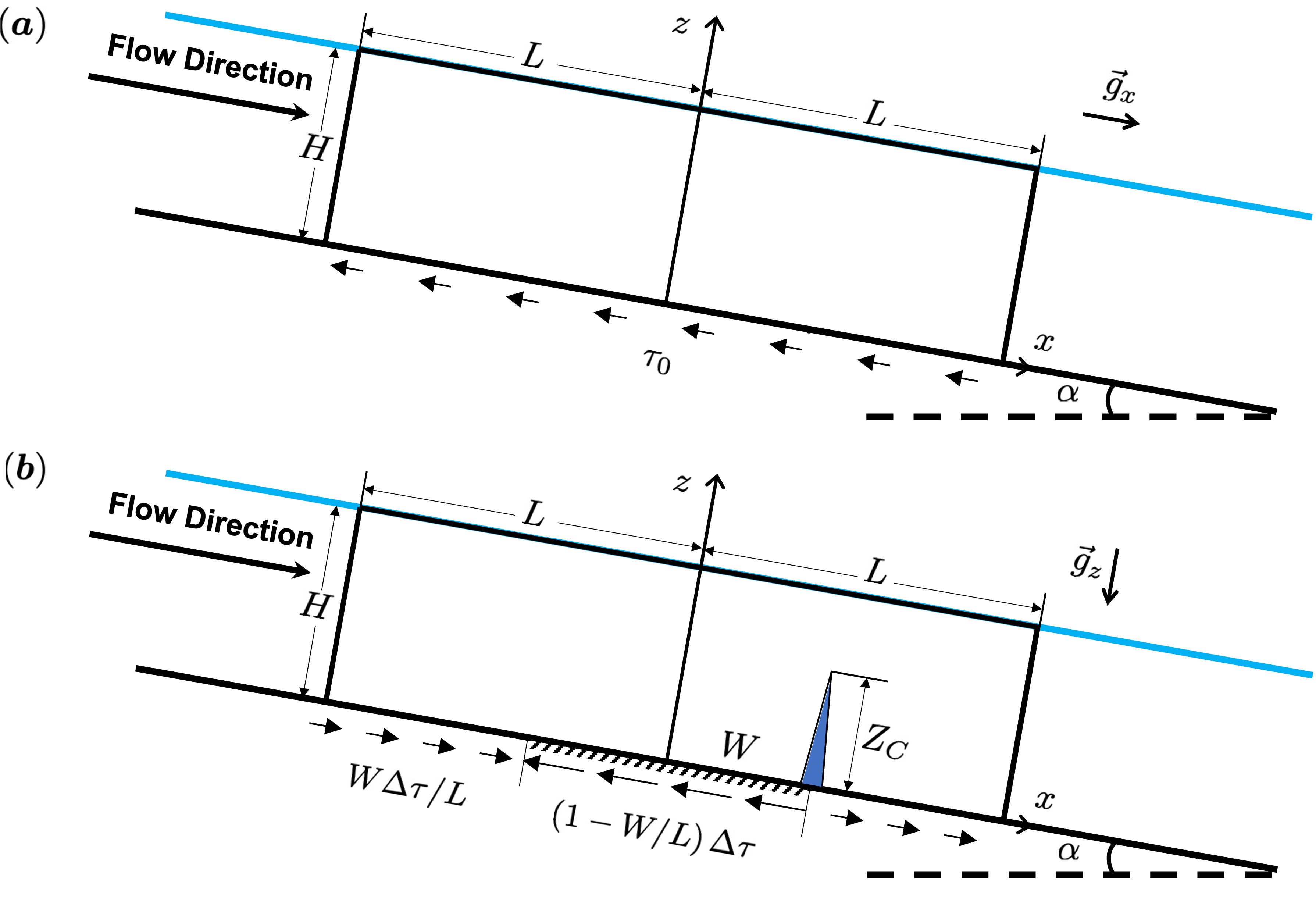}}
    \caption{Schematic diagram of the computational domain, a $2L\times H$ strip extracted from the ice sheet on a slope with angle $\alpha$. The sticky patch is represented by an interval $\left[-W,W\right]$ on the  $x$-axis with excess basal stress $\Delta\tau$. A water-filled crack is initiated on the downstream end of the sticky patch. The crack can be a vertical line with length $Z_{C}$ (as illustrated in the figure), or a curve, depending on the criterion used to determine its propagation. }
    \label{fig:Schematic}
\end{figure}

We model a grounded ice sheet sliding due to gravity, influenced by a sticky patch. Under the assumption of plane strain (i.e., the elastic strain is constrained in the plane of paper), we model the ice sheet with a two-dimensional, infinite elastic strip sliding down a slope with angle $\alpha$. Figure~\ref{fig:Schematic} shows a schematic diagram of the mathematical model, which considers a finite length of the strip that extends across a sticky patch. The total system is decomposed into two different sub-problems, based on the two components of gravity in the tilted coordinate system. The first sub-problem addresses the sliding state, where steady sliding is driven by down-slope component of gravity $\boldsymbol{\vec{g}}_{x}$ and resisted by uniform basal stress (Fig.~\ref{fig:Schematic}a). The second sub-problem is the sticky state, where the sticky patch provides additional basal drag, leading to tensile stress that partially offsets the static compression due to $\boldsymbol{\vec{g}}_{z}$ (Fig.~\ref{fig:Schematic}b). In the first sub-problem, the ice strip with height $H$ and length $2L$ $\left(L\gg H\right)$ is sliding at a constant velocity. A uniform basal shear stress $\tau_{0}$, representing background basal drag, is imposed on $\left(-L,L\right)$. The uniform basal stress $\tau_{0}$ exactly balances the down-slope ($x$) component of gravity, leading to steady sliding with internal deformation (to maintain torque balance), which is consistent with the fact that the ice-sheet motion can be decomposed into basal sliding and internal deformation.

In the sticky state of panel~(b), we focus on the influence of the sticky patch by treating in superposition to the steady, sliding state. To model the stickiness of the sticky patch, an excess shear stress $\Delta\tau$ is imposed on $x\in\left[-W,W\right]$ at $z=0$. Since the down-slope component of gravity is already balanced by $\tau_0$ in panel~(a), this excess traction needs to be balanced. We add an uniform increment of traction $W\Delta\tau/L$ in the opposite direction on $x\in\left[-L,L\right]$ at $z=0$---the whole bottom boundary of the finite ice strip. This extra increment of uniform traction vanishes for $W/L\to 0$ and, for $W/L\ll 1$, has little effect on basal stress estimates around the sticky patch.

In panel~(b), the excess traction on $-W$ to $W$ at $z=0$ creates tension on the downstream side of sticky patch. If that tension, combined with basal water pressure, is large enough to overcome the ice overburden pressure plus the fracture toughness of ice, basal crevassing should occur. In natural ice sheets, the basal drag on the sticky patch $\Delta\tau$ can be much larger than the background basal drag $\tau_{0}$ \citep{sergienko2013regular}. Therefore, in regions close to the sticky patch, the elastic deformation caused by the uniform basal drag is negligible compared with the effect of the patch. From this point forward, we will only focus on the model in panel~(b), which accounts for the effect of the sticky patch .

Based on LEFM theory, the stress intensity factors (SIFs) are calculated and used to predict the maximum penetration of that crack \citep{broek1982elementary}. Physically, stress intensity factors are parameters that describe the stress distribution and magnitude close to the crack tip. A more detailed discussion of stress intensity factors can be found in the following subsections.

\subsection{2D elastic model of basal hydrofracture}

The equation governing the conservation of momentum in ice is
\begin{equation}\label{eq:governing}
    \boldsymbol{\nabla}\cdot\boldsymbol{\sigma}+\rho_i\boldsymbol{g}=\boldsymbol{0},
\end{equation}
where $\boldsymbol{\sigma}$ is the Cauchy stress tensor, $\rho_i$ is the density of ice, and $\boldsymbol{g}$ is the acceleration due to gravity. Here $\boldsymbol{g}=-g\bvec{z}$, where $\bvec{z}$ is the unit vector in the direction of positive-$z$ direction, and $g$ is a constant. With the assumption that slope $\alpha\ll 1$, both side-boundaries of the ice strip are loaded by a depth-dependent compression $\rho_{i}g \left(z-H\right)$ that represents the overburden stress due to the weight of the overlying ice,
\begin{equation}
    \boldsymbol{\sigma}\cdot\boldsymbol{n}=\rho_{i}g \left(z-H\right)\boldsymbol{n}\quad \text{for}\quad x=\pm L,
\end{equation}
where $\boldsymbol{n}$ is the outward-pointing unit normal vector of the domain. The top boundary is assumed to be traction free, neglecting the atmospheric pressure and other traction on the surface,
\begin{equation}\label{eq:bc_top}
    \boldsymbol{\sigma}\cdot\boldsymbol{n}=\boldsymbol{0} \quad \text{for}\quad z=H.
\end{equation}
On the bottom $\left(z=0\right)$, we impose a zero-displacement boundary condition in the $z$-direction and the traction perturbation $\Delta\tau$ in $x$-direction,
\begin{align}\label{eq:bc_bottom_x}
    \boldsymbol{t}\cdot\boldsymbol{\sigma}\cdot\boldsymbol{n} &= \tau\left(x,0\right)=\left\{
    \begin{aligned}
    &-\frac{W}{L}\Delta\tau\quad&\left|x\right|\geq W, \\
    &\left(1-\frac{W}{L}\right)\Delta\tau\quad &\left|x\right|< W,
    \end{aligned}
    \right.\\
    \label{eq:bc_bottom_z}
    \boldsymbol{u}\cdot\boldsymbol{n} &= 0,
\end{align}
where $\Delta\tau$ represents the stress variation caused by a sticky patch and $\boldsymbol{u}$ is the displacement, and $\boldsymbol{t}=-\bvec{x}$ is the unit tangent vector to the boundary. On the bottom, $\boldsymbol{t}$ is in the direction of the negative-x direction. The extra term $-{W}\Delta\tau /L$ is for force balance, as discussed above. Physically, we balance the excess shear stress on the sticky patch using a uniform stress on a much larger area. In our computation, the ratio ${W}/{L}$ is set to be $0.1$, which is small enough to make the effect of the extra term negligible.

Crack walls are loaded by water pressure that is, in general, dependent on the subglacial hydrology. In this study, we assume static basal water pressure and represent subglacial hydrology in terms of the flotation fraction, which is the ratio of basal water pressure to basal overburden pressure,
\begin{equation}\label{def:fl}
    f = \frac{p_{w}}{\rho_{i}g H} = \frac{\rho_{w}g H_{w}}{\rho_{i}g H},
\end{equation}
where $H_{w}$ is the piezometric head measured in borehole experiments as $p_{w}=\rho_{w}g H_{w}$ \citep{harper2010vertical}. The fracture is assumed to initiate at $x=W$, where tensile stress is maximum, and reach a height $Z_{C}$ that is to be determined. Using equation~\eqref{def:fl}, the boundary conditions on the crack walls ($x=W$, $0\le z\le Z_{C}$) can be written in terms of $f$ and $z$,
\begin{equation}\label{eq:bc_crackwall}
    \boldsymbol{\sigma}\cdot\boldsymbol{n}=\left\{
    \begin{aligned}
    \rho_{i}g H\left( \frac{\rho_{w}}{\rho_{i}}\frac{z}{H} - f\right)\boldsymbol{n}\quad &\text{at}\quad z < H_w, \\
    \boldsymbol{0}\quad &\text{at}\quad z \ge H_w.
    \end{aligned}
    \right.
\end{equation}
The transition at $z=H_w$ occurs because water rises to that height in the crevasse.

In formulating the plane-strain elastic constitutive relation, we assume that strain occurs in response to deviations from the overburden stress, as suggested by \cite{cathles2015viscosity}. This approach was used by \cite{lipovsky2020ice} to model rift propagation in floating ice. A perturbation stress tensor $\boldsymbol{T}$ is introduced as the total Cauchy stress tensor $\boldsymbol{\sigma}$ minus the ice overburden stress,
\begin{equation}\label{def:perturbation_sigma}
    \boldsymbol{T}=\boldsymbol{\sigma}+p_{i}\boldsymbol{I},
\end{equation}
in which $p_{i}=\rho_{i}g\left(H-z\right)$ accounts for the ice overburden pressure and $-p_{i}\boldsymbol{I}$ is the ice overburden stress. The elastic constitutive law linearly relates the strain to the perturbation stress as
\begin{equation}\label{eq:constitutive}
    \boldsymbol{T}=\frac{E\nu}{\left(1+\nu\right)\left(1-2\nu\right)} \text{tr}\left(\boldsymbol{\epsilon}\right)\boldsymbol{I}+\frac{E}{1+\nu}\boldsymbol{\epsilon},
\end{equation}
where
\begin{equation}\label{eq:elastic_strain}
    \boldsymbol{\epsilon}(\boldsymbol u)  = \frac{1}{2}\left(
    \nabla \boldsymbol u
    + 
    \nabla \boldsymbol u^T \right).
\end{equation}
To define the constitutive law, two parameters are needed to account for ice properties. Here we use the Young's modulus $E=10$ GPa, and a Poisson's ratio $\nu=0.33$ \citep{van1998fracture}.
Further details associated with the problem description in terms of the perturbation stress are provided in Appendix~A\ref{sec:AppendixA}.

\subsection{Stress Intensity Factors}\label{sec:SIF}
In LEFM theory, stress intensity factors $K_{\RNum{1}}$, $K_{\RNum{2}}$ are used to describe the additional stress near the tip of a crack that is due to the presence of the crack \citep{broek1982elementary},
\begin{gather}
    \sigma_{i j}^{\RNum{1}}(r,\theta)=\frac{K_{\RNum{1}}}{\sqrt{2\pi r}}f_{i j}^{\RNum{1}}\left(\theta\right)+\textit{O}\left(r^{\frac{1}{2}}\right),\label{eq:SIF_1}\\
    \sigma_{i j}^{\RNum{2}}(r,\theta)=\frac{K_{\RNum{2}}}{\sqrt{2\pi r}}f_{i j}^{\RNum{2}}\left(\theta\right)+\textit{O}\left(r^{\frac{1}{2}}\right),\label{eq:SIF_2}
\end{gather}
where $\RNum{1}$, $\RNum{2}$ represent two in-plane modes of cracks (opening and shearing, respectively), $r,\theta$ are coordinates in a local plane-polar coordinate system with origin at the crack tip, and $f_{ij}^{\RNum{1}}$, $f_{ij}^{\RNum{2}}$ are tensor-valued functions describing the angular dependence of the excess stress. The $O\left(r^{1/2}\right)$ terms are related to far-field stress and are omitted when $r$ is small. 

The predicted extent of the crack that is in equilibrium with the total stress field, including its direction (if not constrained to be vertical), is determined by the criterion of crack propagation. The broadly accepted $G$-criterion states that cracks grow in the direction along which the maximum potential energy is released \citep{broek1982elementary}. In the context of this study, the potential energy is the strain energy stored in the elastic ice sheet. The elastic energy released when a crack extends is measured by the release rate $G$, defined is the elastic energy released per unit of crack extension. The crack extends when $G$ is greater than or equal to $G_{C}$, the energy required for crack growth. Equilibrium is attained when this condition is no longer met and the crack ceases to propagate.  

$G$ is related to the stress intensity factors by
\begin{equation}\label{eq:G-criterion}
    G=G_{\RNum{1}}+G_{\RNum{2}}=\frac{1-\nu^2}{E}\left(K_{\RNum{1}}^2+K_{\RNum{2}}^2\right),
\end{equation}
where $G_{\RNum{1}}$ and $G_{\RNum{2}}$ denote the energy release rate for mode-I and mode-II components, with $G_{C}$ given by
\begin{equation}
    G_{C}=\frac{1-\nu^2}{E}K_{C}^2,
\end{equation}
where $K_C$ is the fracture toughness, a material property measured in experiments. The $G$-criterion $\left(G>G_{C}\right)$ can be expressed in terms of the stress intensity factors as $K=\sqrt{K_{\RNum{1}}^2+K_{\RNum{2}}^2}>K_{C}$. Therefore, extension of a crack occurs when the total stress intensity factor exceeds fracture toughness $K_C$. We assume that $K_{C}=100~\text{kPa m}^{1/2}$ \citep{rist1996experimental}, which is an estimate widely used in ice-fracture problems. 

In the results section, we first assume a vertical, pure mode-I crack and employ a simple criterion for its propagation. In particular, we neglect the mode-II component by assuming that $K_{\RNum{2}}=0$. Thus the criterion for crack extension reduces to $K_{\RNum{1}}>K_{C}$; if satisfied, the crack can grow vertically in length. Later we reconsider mixed-mode crack growth with $K_{\RNum{2}}\ne0$ and assess the difference associated with the different criteria.

\subsection{Non-dimensionalisation}\label{sec:nondimensionalisation}
In section Results, we focus on non-dimensional problems and denote the non-dimensional version of a variable using the same symbol appended with a prime $^{\prime}$.
We choose to scale length (and displacement) by $H$, and the stress by $\rho_i g H$.
Hence, the conservation of momentum (Eqs.~\eqref{eq:governing}--\eqref{eq:bc_crackwall}) and stress intensity factors are non-dimensionalised by the following scales:

\begin{align}
    \boldsymbol x^{\prime}=\frac{\boldsymbol x}{H},\quad 
    \boldsymbol u^{\prime}=\frac{\boldsymbol u}{H},\quad     
    \boldsymbol{\sigma}^{\prime}=\frac{\boldsymbol{\sigma}}{\rho_{i}g H},\quad \Delta\tau^{\prime}=\frac{\Delta\tau}{\rho_i g H},\quad K_{\RNum{1},\RNum{2},C}^{\prime}=\frac{K_{\RNum{1},\RNum{2},C}}{\rho_i g H^{3/2}}.
\end{align}
Beside the flotation fraction representing basal water pressure, there are two important, non-dimensional parameters that control fracturing, $W^{\prime} = W/H$ and $\Delta\tau^{\prime}$, which represent the size and excess shear stress of the sticky patch, respectively. The non-dimensional stress intensity factors and non-dimensional fracture toughness are used in the fracture criterion. Note that although the dimensional fracture toughness (measured by experiments) is purely empirical, the non-dimensional fracture toughness is scaled by $\rho_{i}g H^{3/2}$, giving it a dependence on the ice-sheet thickness.

\subsection{Numerical Implementation}\label{sec:numerical}
For the vertical-line-crack problem, the governing equations are solved using an open-source finite element (FE) software library, FEniCS \citep{logg2010dolfin,logg2012automated,LangtangenLogg2017}, with meshes generated by Gmsh \citep{geuzaine2009gmsh}. In the discretised problem, the basal crevasse is represented by a straight, triangular notch in the mesh, perpendicular to the bottom boundary. The mesh is locally refined near the tip of the notch and the bottom boundary. We use triangular elements in the mesh, with element size varying from $5\times10^{-3}$ times the crack lengths near the tip to $0.2$ ice thicknesses (away from the tip). For an $1000$m-thick ice sheet with a $500$m-long basal crevasse, the element sizes are $2.5$~m near the tip and $200$~m away from the tip. SIFs are calculated by Displacement Correlation Method (DCM) \citep{chan1970finite, banks1986comparison} with Richardson extrapolation \citep{guinea2000ki}. The code is benchmarked by comparison with weight function from \cite{jimenez2018evaluation} (see Appendix~B\ref{sec:AppendixB}).

The FEniCS code relies on meshing the interior of the elastic domain, and requires local mesh refinement at the fracture tip to resolve the stress singularity there. To perform simulations of fracture growth where the path is not specified \textit{a priori}, this method would require repeated remeshing.  Dynamic remeshing is feasible with specialised software but is unreliable and incurs a significant computational cost. To avoid such issues and to provide an additional means to verify the numerical results, we use a separate code for consideration of curved fracture trajectories.  This code employs the Displacement Discontinuity Method (DDM) \citep{crouch1982boundary}, a scheme that is based on the Boundary Element Method (BEM). 

In the DDM approach, only boundaries subject to given conditions must be meshed.  The interior of the domain is assumed to be a uniform, linear elastic material that deforms according to Green's functions forced by the boundary facets. This method avoids reliance on meshing software. To extend the fracture at a given step, a new straight-line segment is connected to the former tip of the fracture. The orientation of this new segment is defined by the optimal direction of fracture growth. The new segment must have an appropriately small length, the choice of which defines the resolution of the model. At each growth step, we test for multiple orientations of growth and choose that with the highest strain-energy release rate, as defined by the $G$-criterion \citep{dahm2000numerical}. The BEM code used here is developed by \cite{davis2017new}. As with the FE models, in the DDM models we subject the base of the glacier domain to the condition of no vertical displacement. 

\section{Results}\label{sec:results}

Results are presented in two parts.  The first is for vertical fractures and uses the FE model; the second is for mixed-mode, curving fractures and uses the DDM model.

\subsection{Mode-I fracture growth}
\begin{figure}
    \centering{\includegraphics[width=0.9\textwidth]{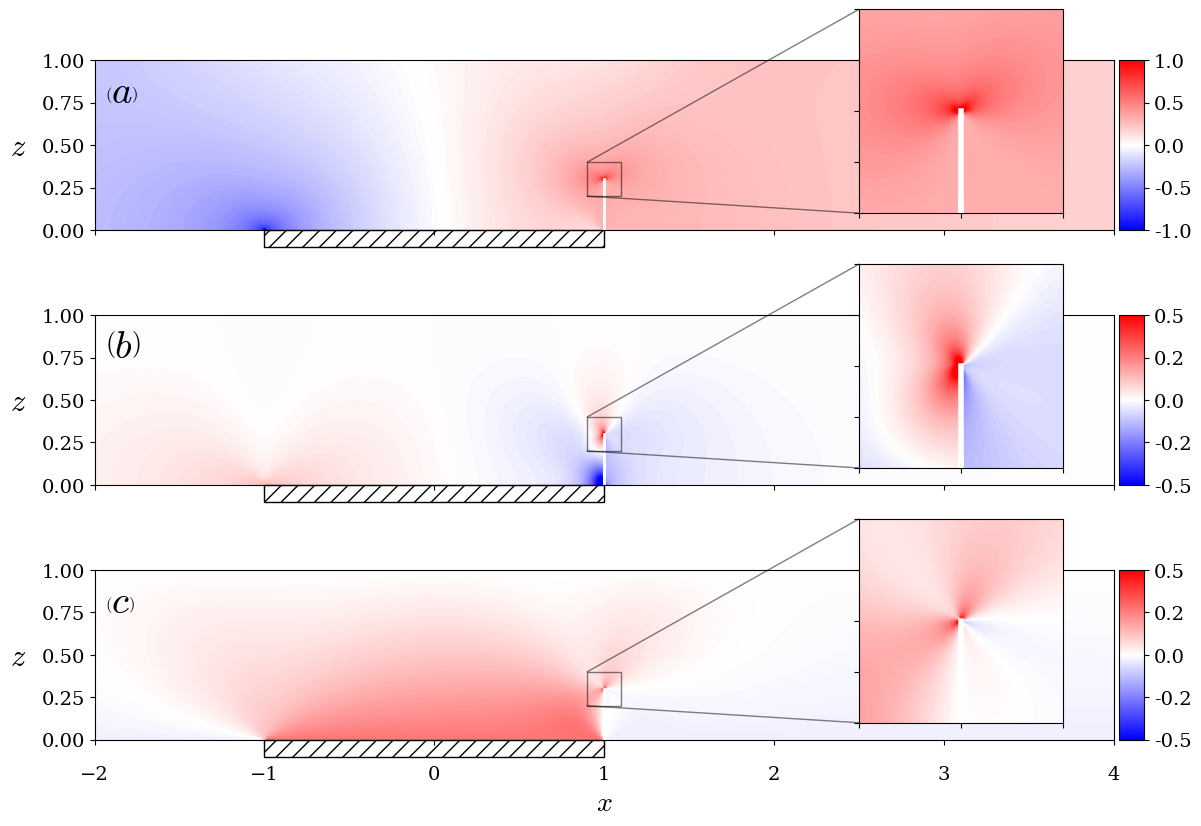}}
    \caption{Perturbation stress in a cracked ice-sheet with $W^{\prime}=1$, $\Delta\tau^{\prime}=0.3$, and $Z_C^{\prime}=0.3$. Panels show \textbf{(a)}~$T_{xx}^{\prime}$, \textbf{(b)}~$T_{zz}^{\prime}$, and \textbf{(c)}~$T_{xz}^{\prime}$. Tension is defined as having a positive sign. In the zoom-in box a white rectangle is added to highlight the crack, which is actually much narrower, making it difficult to see otherwise.}
    \label{fig:sigma_perturbation_1}
\end{figure}

We consider a vertical, mode-I fracture as part of a reference case where the sticky patch has a width two times the ice thickness ($W'=1$).  The non-dimensional excess shear stress is $\Delta\tau^{\prime}=0.3$, which means the dimensional excess shear stress is $0.3\rho_{i} gH$. The reference flotation fraction is $f=0.7$. Figure~\ref{fig:sigma_perturbation_1} shows the three components of the perturbation stress that arise due to the excess stress on the boundary. Among these, we are most interested in $T_{xx}^{\prime}$ and $T_{xz}^{\prime}$, which are related to the fracture propagation. As shown in panel~(a), there is horizontal tension  concentrated on the downstream (right) end of the sticky patch and compression concentrated on the upstream (left) end. In panel~(b), the vertical normal stress $T_{zz}^{\prime}$ is also concentrated near the ends of the sticky patch and the crack tip. This is a consequence of the no-vertical-displacement condition \eqref{eq:bc_bottom_z} imposed on the bottom boundary. Panel~(c) shows the pattern of shear stress, which is localised to a region around the sticky patch.

\begin{figure}
    \centering{\includegraphics[width=0.4\textwidth]{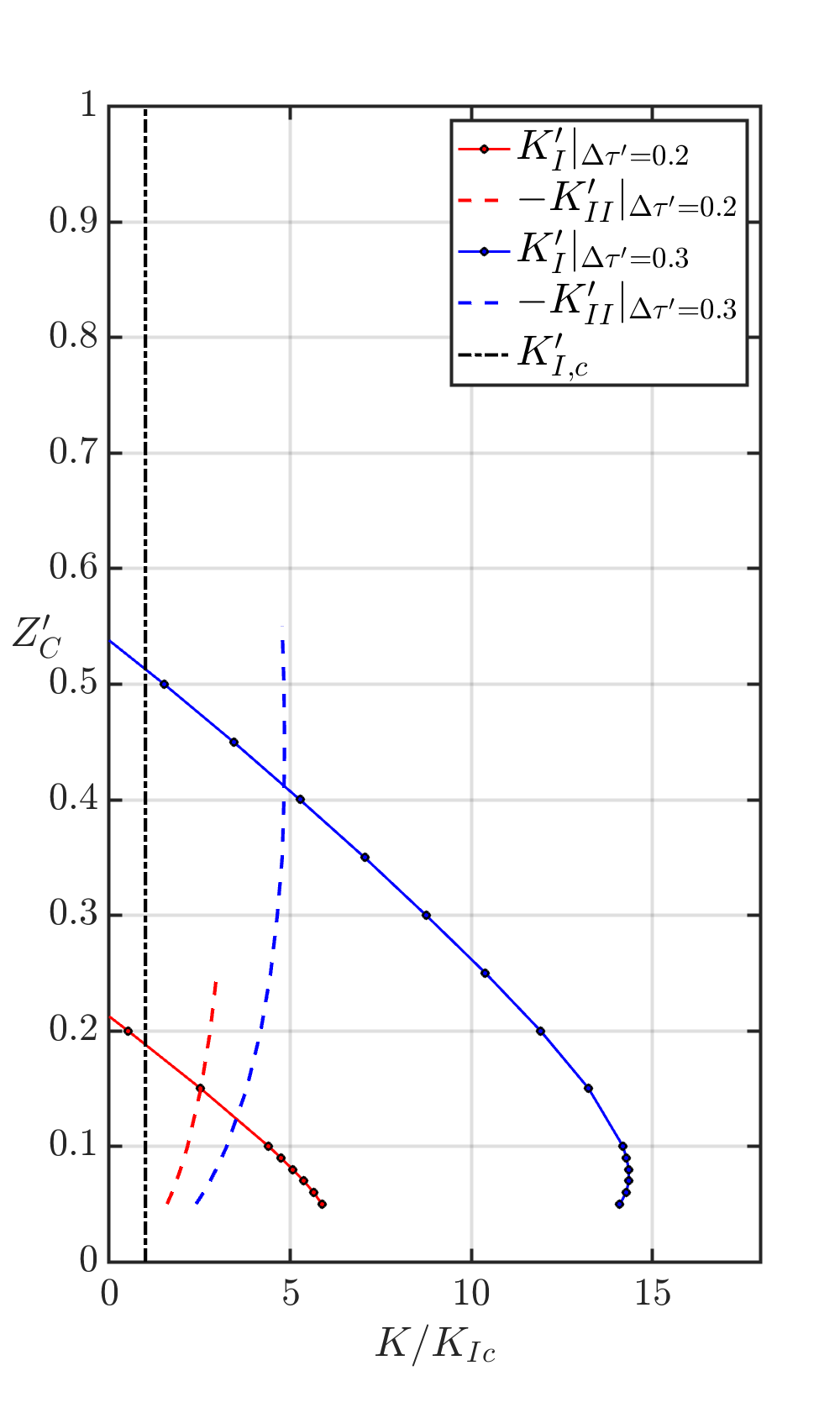}}
    \caption{Stress intensity factors versus crack length with $W^{\prime}=1$.  Two cases with different $\Delta\tau^{\prime}$ are considered: $\Delta\tau^{\prime}=0.2$ (red) and $\Delta\tau^{\prime}=0.3$ (blue). The black dash-dotted line represents non-dimensional fracture toughness $K^{\prime}_{\RNum{1},c}={K_{\RNum{1},c}}/\left({\rho_{i}g H^{3/2}}\right)$, where $H$=100~m.}
    \label{fig:K1_lambda}
\end{figure}

 We are interested in the maximum stable crack length permitted by the fracture criterion, and how this depends on ice-sheet thickness and sticky-patch width. To investigate, we extend our calculation to cases with varying crack length $Z_{C}^{\prime}$ and assume that the criterion for crack growth is $K_{\RNum{1}}^{\prime}>K_{c}^{\prime}$ \citep{van1998fracture}. Figure~\ref{fig:K1_lambda} shows the two stress intensity factors as functions of $Z_{C}^{\prime}$ near the sticky patch for the reference model ($W^{\prime}=1$, $\Delta\tau^{\prime}=0.3$). With the crack length $Z_{C}^{\prime}$ increasing, $K_{\RNum{1}}^{\prime}$ increases to its maximum due to tension near the sticky patch, then drops to negative values due to overburden pressure. The maximum crack length is the value of $Z_{C}^{\prime}$ such that $K_{\RNum{1}}^{\prime}=K_{c}^{\prime}$ (i.e., the intersection of the $K_{\RNum{1}}^{\prime}$ curve and vertical $K_{c}^{\prime}$ line in Figure~\ref{fig:K1_lambda}). For the crack with this value of $Z_{C}^{\prime}$ to be in a stable equilibrium, the crack should be resistive to perturbations. If we add a positive perturbation $\Delta Z_{C}^{\prime}>0$ to the crack length $Z_{C}^{\prime}$, the perturbed crack length should return to the stable state, which means $K_{\RNum{1}}^{\prime}\left(Z_{C}^{\prime}+\Delta Z_{C}^{\prime}\right)<K_{\RNum{1},c}^{\prime}$. Thus, the condition for a stable crack is ${d K_{\RNum{1}}}^{\prime}/{d Z_{C}^{\prime}}<0$. 
 
\begin{figure*}
    \centering{\includegraphics[width=1\textwidth]{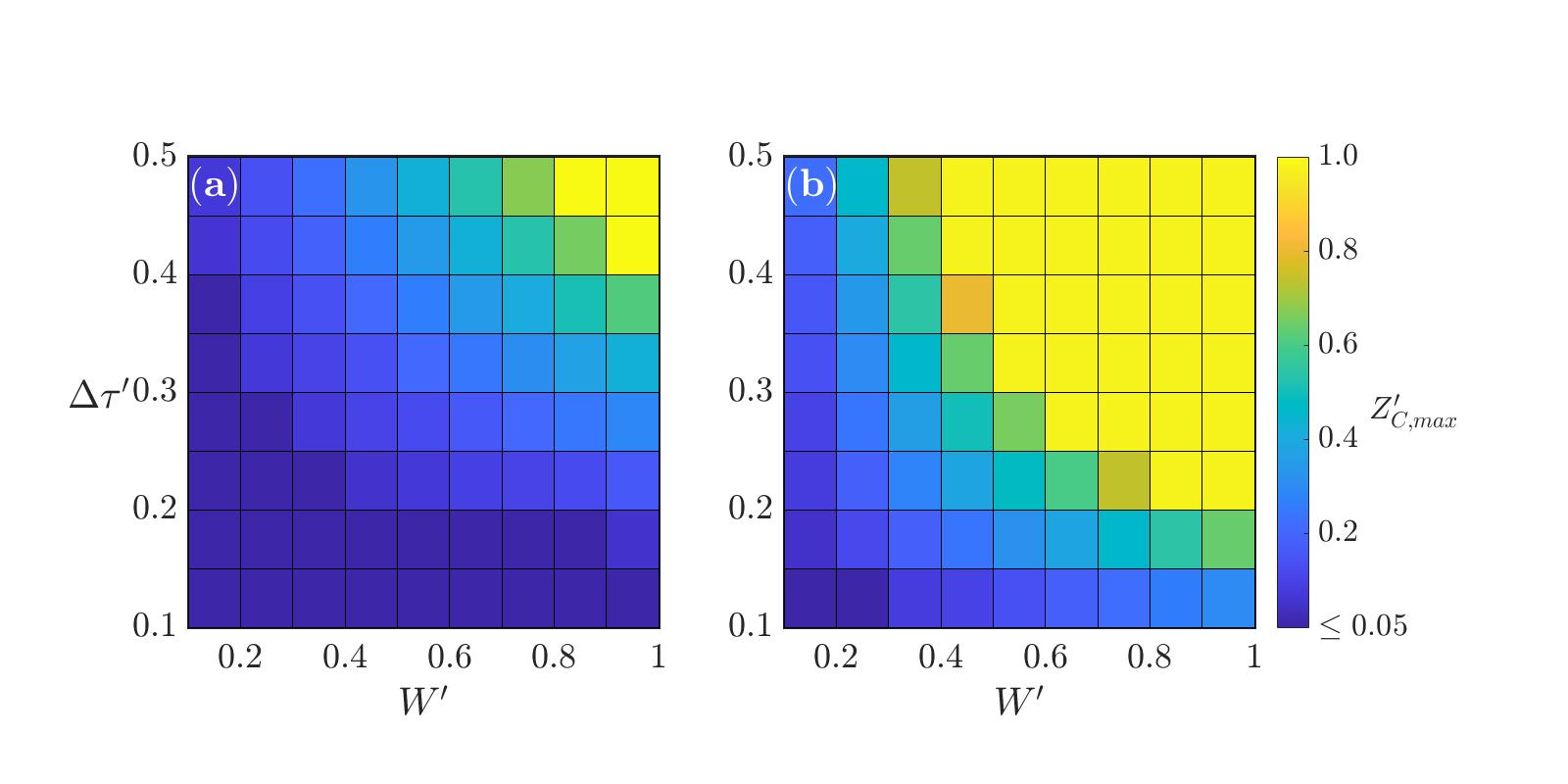}}
    \caption{Maximum crack length $Z_{C,max}^{\prime}$ as a function of $W^{\prime}$ and $\Delta\tau^{\prime}$. Note that in the criterion, the non-dimensional fracture toughness depends on the ice thickness $H$. Here we set $H=100\ \text{m}$. For larger thicknesses, $K_{C}^{\prime}$ gets smaller, leading to longer cracks. Two cases with different $f$ are considered: \textbf{(a)}~$f=0.7$, \textbf{(b)}~$f=0.9$. When the crack length $Z_{C}^{\prime}=1$, the crack dissects the full ice-sheet thickness.}
    \label{fig:dwmu}
\end{figure*}

Figure~\ref{fig:dwmu} shows the maximum stable crack length $Z_{C,max}^{\prime}$ as a function of dimensionless problem parameters $\Delta\tau^{\prime}$ and $W^{\prime}$ for an ice thickness of $100$~m. In panel~(a), flotation fraction is $f=0.7$. The maximum crack length increases monotonically with $W^{\prime}$ or $\Delta\tau^{\prime}$. For sticky patches with $W^{\prime}=1$, a minimum $\Delta\tau^{\prime} \approx 0.2$ is required to initiate cracks. Panel~(b) shows another case where the ice is closer to flotation (i.e., $f=0.9$). Here, fracture can be initiated by smaller values of $W^{\prime}$ and $\Delta\tau^{\prime}$, such as $W^{\prime}=0.5$, $\Delta\tau^{\prime}=0.1$. The sticky patch creates a localised horizontal tension that promotes the growth of the vertical line crack, as shown by the reference case in Figure~\ref{fig:sigma_perturbation_1}. Evidently, this effect is sensitive to the non-dimensional parameters $W^{\prime}$ and $\Delta\tau^{\prime}$.
    
\subsection{Mixed-mode fracture growth}
A limitation of the models above is that they assume basal crevasses are mode-I cracks that only propagate vertically and, to simplify the calculation, neglect the mode-II component. This approach is consistent with previous LEFM models of basal crevasses that are pure hydrofractures \citep{van1998fracture}. However, with excess shear stress arising from the sticky patch, basal crevasses are expected to have both mode-I and mode-II contributions. Relaxing our assumption that the crack propagates vertically, we now consider the curved, quasi-static path of fractures using the BEM implementation. Note that for curved fracture paths, we use the $G$-criterion for fracture growth: the fracture stops when either $G<G_c$ or when there is closure of fracture walls just behind the fracture's tip ($K_I\le0$).

\begin{figure}
    \centering{\includegraphics[width=1.0\textwidth]{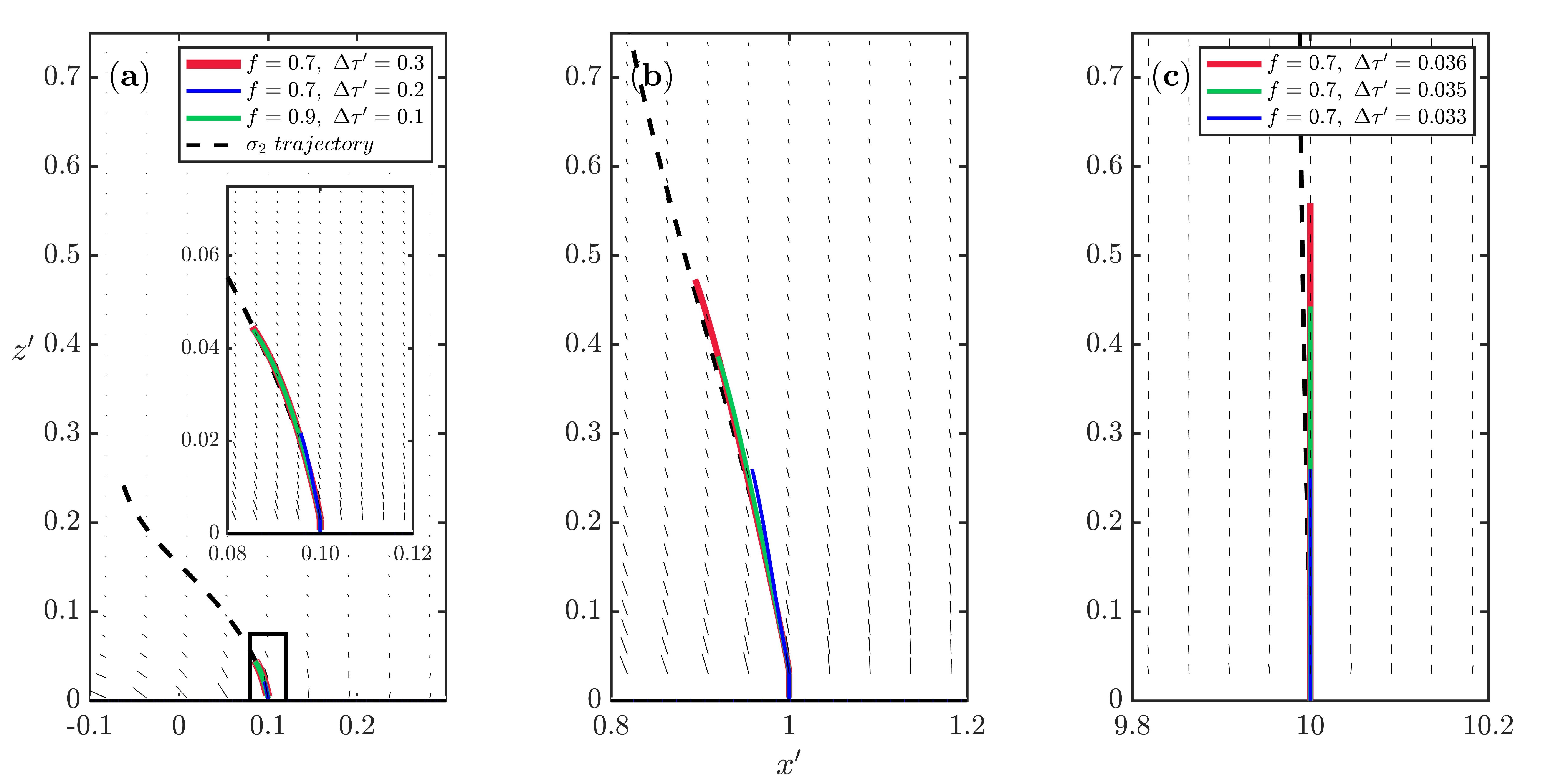}}
    \caption{Fracture paths calculated under different excess basal shear stress $\Delta\tau^{\prime}$ and flotation fraction $f$, with three values of $W^{\prime}$. The background vector field indicates the local direction of the maximum compressive stress, with the vector length scaled by the deviatoric stress $\left|{T_{1}^{\prime}-T_{2}^{\prime}}\right|$, where $T_{1}^{\prime}$ and $T_{2}^{\prime}$ are the local principal stresses calculated from $\boldsymbol{T}^{\prime}$. In panels~(a) and (b), the fracture paths (colors) are calculated by BEM, with the principal stress trajectory represented by the dashed curve. In panel~(c), since the principal stress trajectory is nearly a vertical line, we no longer conduct the BEM simulation and just assume purely vertical fracture paths. Thus, the maximum crack length is determined as we have done in Figure~\ref{fig:K1_lambda}. An ice thickness $H=10^{3}$~m is used to scale the fracture toughness. \textbf{(a)} The growth path when $W^{\prime}=0.1$, the colors represent three different combinations of $\Delta\tau^{\prime}$ and $f$. \textbf{(b)} The growth path when $W^{\prime}=1$, where the legend is the same as panel~(a). \textbf{(c)} The growth path when $W^{\prime}=10$. Note the fracture propagation is stopped when either $G<G_c$ or when there is contact of fracture walls at the fracture tip. In the case of fracture-wall contact, growth at the tip will be in the form of a wing crack.}
    \label{fig:fracture_BEM}
\end{figure}

Figure~\ref{fig:fracture_BEM} shows the mixed-mode fracture paths with different values of dimensionless excess shear stress $\Delta\tau^{\prime}$ and flotation fraction $f$.  Each of the three panels shows results for a different value of $W^{\prime}$. In panel~(a), we consider $W^{\prime}=0.1$, a sticky patch whose width is one fifth of the ice thickness. There is a zoom-in plot of the fracture paths in the black box. The blue and red curves are quasi-static paths where flotation fraction $f=0.7$ and excess shear stress is $0.2 \rho_{i}gH$ (blue) and $0.3 \rho_{i}gH$ (red). The green curve, which almost overlaps the red curve, is the path with a higher flotation fraction $f=0.9$ and lower excess shear stress $\Delta\tau=0.1\rho_{i}gH$. The paths deviate from the vertical path assumed in models above. In particular, they incline upstream under the influence of the shear stress. Comparing the three curves in panel~(a), we find that $\Delta\tau^{\prime}$ and $f$ have little effect on the direction of the paths. 

In panel~(b), with $W^{\prime}=1$ (sticky patch width twice the ice thickness; an order of magnitude larger than in panel~(a)), the fractures align closer to the vertical and extend further into the ice sheet.

Panel~(c) shows the fracture paths when $W^{\prime}=10$. In this case the sticky patch is 20 times wider than the ice thickness. The magnitude of the excess shear stress $\Delta\tau^{\prime}$ required for fracture initiation is reduced to about $0.035$. Meanwhile, since $T_{xx}^{\prime}\gg T_{zz}^{\prime}$, the principal stress trajectories are nearly vertical lines, as indicated by the background stress field. Thus, fracture paths can be approximated by vertical lines, as we did previously. The crack length becomes very sensitive to the magnitude of $\Delta\tau^{\prime}$, since a small perturbation to $\Delta\tau^{\prime}$ (from $0.033$ to $0.036$) causes a large variation of the crack length.

It is possible to predict the trajectory of the curved cracks without solving the BEM model, using only stresses computed in an uncracked domain. This may be computationally convenient in combination with Stokes-flow models of ice sheets \citep{krug2014combining, yu2017iceberg}. As in the models discussed above, the excess shear stress due to the sticky patch is imposed as a boundary condition on the uncracked domain. We then calculate the perturbation stress in the uncracked domain and plot the principal stress trajectories as the dashed curves in Figure~\ref{fig:fracture_BEM}. We use the perturbation stress due to the excess shear only, because the overburden stress and water pressure are isotropic and do not contribute to the orientation of the principal stresses. Comparison between these trajectories and the fracture paths predicted by the BEM shows that the former are accurate approximations of fracture paths under the conditions considered here. 

If there are no deviatoric stress from sources other than the sticky patch, the magnitude of $\Delta\tau^{\prime}$ doesn't contribute to the direction of the fracture path. Instead, the direction of trajectories depends only on the ratio $W^{\prime}$. This is consistent with the results predicted by the BEM and indicates that for real basal crevasses affected by sticky patches, the direction of the fracture is predominantly controlled by the relative size of the sticky patch.

\section{Discussion}
We have developed and analysed a model of basal crevassing associated with sticky patches at the bed of an elastic glacier or ice sheet. Our model, based on LEFM theory, evaluates the role of shear-stress variations and makes predictions of crack lengths and trajectories. As shown above, the growth of such basal cracks depends on the flotation fraction $f$, the non-dimensional size of the sticky patch $W^{\prime}$ and the non-dimensional stress variation $\Delta\tau^{\prime}$.

\subsection{Potential applications to real ice sheets}
 
For real ice streams, basal shear stress patterns are difficult to observe directly. In several cases, they have been estimated by inversion of surface data under specific assumptions. \cite{sergienko2013regular} showed that for a region of the Antarctic ice sheet with large ice thickness $\left(H\approx 10^{3}\ \text{m}\right)$, the basal shear stress has rib-like patterns of variation. The width of such sticky patches varies from one ice thicknesses ($W'=0.5$) to $10$ ice thicknesses ($W'=5$). The excess shear stress ($200$ to $300$~kPa) estimated for these locations is small compared with the overburden pressure ($\Delta\tau^{\prime}\sim 0.03$). In this case, the effect of the sticky patch depends on its size and local water pressure, as discussed in the Results section. For small patches ($W^{\prime}\sim 1$), such excess shear stress only affects the direction of fracture propagation, as shown in Figure~\ref{fig:fracture_BEM}. Basal crevassing occurs when the flotation fraction $f$ reaches $1$, which is consistent with the conclusions of \cite{van1998fracture}. For larger patches with $W^{\prime}\sim 10$, basal crevasses would initiate on the downstream end at a lower water pressure $f\sim 0.7$.

For alpine glaciers with an ice thickness of order 100~m, basal shear stress was found to be in the range of $0$ to $200$~kPa \citep{braedstrup2016basal}. In the non-dimensional parameter space, $\Delta\tau^{\prime}$ is of order $0.1$, larger than in Antarctic settings. Moreover, $W^{\prime}$ is of the order of $1$ to $10$, which is similar to the that of sticky patches in Antarctic ice sheets. Thus, we predict that sticky patches play a more important role in determining the length and direction of basal crevasses in  alpine glaciers. For further investigations of the basal hydrofractures around sticky patches, we need to go beyond the inversion results, understand the specific causes of the sticky patches, and include the essential physics in our model.

In natural glaciers and ice sheets, there are many factors that might create sticky patches, including bedrock bumps, till-free areas, well-drained tills and basal freeze-on. All of these would lead to localised, high basal friction \citep{stokes2007ice}. An important friction phenomenon is the stick-slip motion of ice, detected in Whillans Ice Stream (WIS) in West Antarctica. \cite{wiens2008simultaneous} investigated the WIS stick-slip motion and related it to a sticky patch on the bed. Furthermore, \cite{sergienko2009stick} argued that WIS can be considered as a typical stick-slip system controlled by the basal friction, where the sticky spot nucleates the stick-slip cycle. Therefore, for sticky spots associated with stick-slip ice motion, we can estimate the stress variation from parameters in relevant friction experiments \citep{mccarthy2017temperature,lipovsky2019glacier,zoet2020application}, rather than from inversion from surface data based on viscous rheology. If we interpret $\Delta\tau$ in terms of friction coefficient variation $\Delta\mu$, we find that
\begin{equation}\label{eq:effective}
    \Delta\tau^{\prime}=\Delta\mu N^{\prime},
\end{equation}
where $N^{\prime}=1-f$ is the dimensionless effective normal stress. The basal stress variation is controlled by both friction coefficient variation $\Delta\mu$ and the flotation fraction $f$. The magnitude of $\Delta\mu$ during the stick-slip motion of ice can be measured experimentally. By keeping a constant normal stress of $N=500$~kPa between the ice sample and the bedrock asperity, \cite{zoet2020application} found that the friction-coefficient decrease during stick-slip motion is between $0.1$ and $0.4$. We assume that the measurements of $\Delta\mu$ also applies to the sticky patch discussed in our model.

Well-drained till could serve as a sticky patch. For a well-drained till surrounded by a water saturated layer \citep{stokes2007ice}, the water pressure on the till would be smaller than the surroundings. For such a till, we assume that the excess friction coefficient $\Delta\mu=0.4$ and flotation fraction $f=0.7$. Then, according to \eqref{eq:effective}, the excess basal shear stress is $\Delta\tau^{\prime} \sim 0.1$, which would lead to a tensile-stress concentration on the downstream end of the patch. Meanwhile, in the surroundings, the local subglacial water pressure is expected to be higher ($f\ge 0.7$). For the case considered above, when $f$ reaches $0.9$, basal crevassing is likely to occur.

\subsection{Thermal implications of basal crevasses}
The temperature structure of ice has important effects on ice dynamics. Using high-vertical-resolution sensing, \cite{law2021thermodynamics} reported spatial heterogeneity of englacial ice temperature and deformation.  They found a basal temperate ice zone with thickness that varies from 5~m to 73~m at two locations separated by only $9~\text{km}$ at Store Glacier, an outlet glacier of the Greenland Ice Sheet. Their study indicates spatially varying basal thermal conditions over distances of a few ice thicknesses. Injection of water-filled basal crevasses can locally modify the thermal profile of the ice sheet \citep{luckman2012basal} and is a potential explanation of the spatially-varying temperate ice layer. The thermal structure of basal crevassing, which is similar to that of dykes in rock \citep{daniels2014thermal}, has been modelled in several studies \citep{jarvis1974thermal,mcdowell2021cooling}. Their approach recognises that water-filled basal crevasses in sub-temperate ice propagate on a short timescale, followed by rapid refreezing of water inside the crack.  \cite{mcdowell2021cooling} model refreezing of a basal crevasse as an instantaneous heat source in one-dimensional heat-conduction system, using the analytical solution of \cite{carslaw1959conduction}. We use the same analytical solutions for a two-dimensional thermal structure of basal crevasse. Details of this calculation are provided in Appendix C\ref{sec:AppendixC}. 

To estimate the heat released by refreezing, we return to the dimensional problem and assume a static, water-filled vertical crack with crack length $Z_{C}=50$~m in a $100$ m-thick, subtemperate ice sheet. The crack width $w$ is assumed to be $10$~cm, uniformly along the crack. The water inside the crack instantaneously refreezes at $t=0$, releasing an amount of heat $q_{i}$ per unit length per unit depth into the page (i.e., in the direction normal to the crack). The heat is estimated as
\begin{equation}\label{eq:heat}
    q_{i}=\rho_{i} L w= 3\times 10^{7}\ \text{J}\, \text{m}^{-2},
\end{equation}
where $\rho_{i}w$ is the mass of water per unit area of the fracture and $L$ is the latent heat of solidification. Mathematically, the refreezing process is assumed to be an instantaneous source at $t=0$. The temperature rise cause by refreezing, $\Delta T$, is held at $0$~K on the surface boundary, the bottom boundary and at the limit of $x\rightarrow \pm\infty$. Temperature will asymptotically decay to zero after a long cooling process.

\begin{figure}
    \centering{\includegraphics[width=0.75\textwidth]{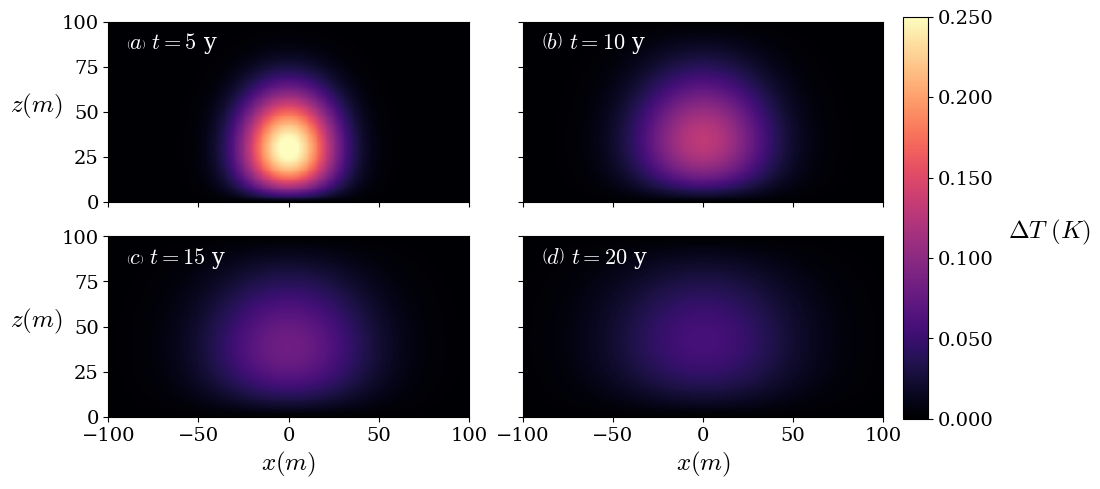}}
    \caption{Perturbation of a single basal crevasse on background temperature field. $t=0$ is the time of crevasse opening and refreezing. \textbf{(a)}~$t=5$ years. \textbf{(b)}~$t=10$ years. \textbf{(c)}~$t=15$ years. \textbf{(d)}~$t=20$ years.}
    \label{fig:Heat Diffusion}
\end{figure}

Figure~\ref{fig:Heat Diffusion} shows the perturbation in temperature due to the refreezing of water in a single basal crevasse over twenty years. The surrounding ice undergoes a rapid warming at $t=0$, followed by a long cooling period until the temperature drops back to the background state $\Delta T=0$ \citep{mcdowell2021cooling}. Panel~(a) shows that after 5 years of diffusion, the temperature perturbation due to refreezing is localised around the crevasse and decreases sharply within $\pm30$~m. Panels~(b) and (c) show the temperature perturbation after 10, 15 and 20 years. After 20 years, $\Delta T$ drops back below $0.05$~K and the system is again close to the background state. 

If a sticky patch is fixed on the bedrock beneath a sliding ice sheet, the patch could generate a series of basal crevasses, as shown in Figure~\ref{fig:heat_series}. Thus, we next consider a case in which the spacing between these crevasses is equal to the half-width of the sticky patch $W=100~\text{m}$. For these equally spaced basal crevasses, the temperature field is a linear superposition of the effect of each crevasse. In the mathematical model, the basal crevasse is initiated on the downstream end of the sticky patch and subsequently advected by sliding. Details are provided in Appendix C\ref{sec:AppendixC}. 

\begin{figure}
\centering{\includegraphics[width=1\textwidth]{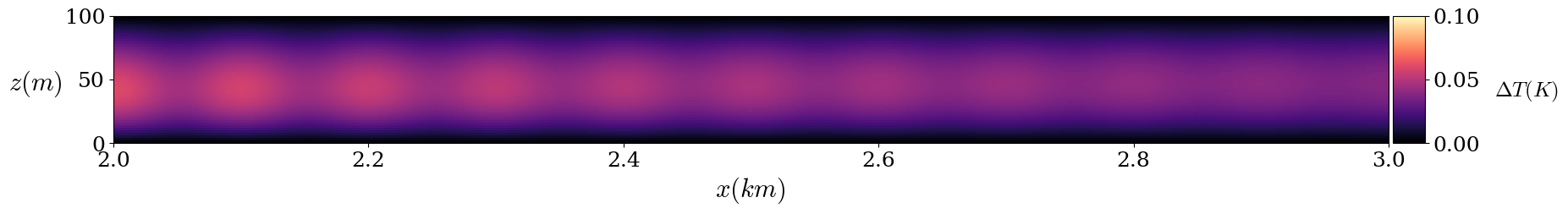}}
    \caption{Thermal effect of a series of basal crevasses ice that is $2$~km to $3$~km downstream from the sticky patch. After a long time of diffusion, the localised temperature field smooths out.}
    \label{fig:Heat Diffusion_series}
\end{figure}

The thermal effect of a series of basal crevasses is shown in Figure~\ref{fig:Heat Diffusion_series}. The downstream perturbation will gradually smooth out after decades of cooling. The stable pattern of temperature rise depends on the volume of water inside the crack. The BEM simulations show that the width of basal crevasses is of order $0.1$~m, which is much smaller than width of surface crevasses from observations. Here we simply assume that the crack width is $0.1$~m. The initial, localised temperature rise will decay rapidly as it smooths out. Refreezing in basal crevasses is a possible factor influencing the temperature profile of basal ice, leading to localised heating around the relic basal crevasses, followed by cooling back toward a steady state \citep{mcdowell2021cooling}.

Alternatively, if the basal fracture is embedded in temperate ice and hence doesn't freeze rapidly, it becomes a persistent mechanical perturbation to the ice sheet. A series of such basal crevasses will be carried downstream to the grounding line. If an ice shelf extends seaward from the grounding line, the basal crevasses will weaken the shelf to lateral shear stresses associated with buttressing and vertical shear associated with calving \citep{bassis2015evolution}. The recent numerical model developed by \cite{berg2022crevasse} suggests that the advection of crevasses could increase the calving rate and promote glacier retreat.

\subsection{Limitations of the Model}
The model presented has three significant limitations. First, it is based on equilibrium equations of elasticity. The ice sheet is assumed to be an isotropic, elastic body without any internal viscous deformation that is commonly computed by a full-Stokes model. Therefore, our model only accounts for the physics of fracturing on a short timescale in which ice is dominated by elastic rheology. This approach might miss important interactions between the sticky patch and viscous deformation that would modify the stress field. Thus, a viscoelastic rheology is required to study basal crevasses on a longer timescale. Moreover, it is important to include the spatial and temporal variation of subglacial hydrology \citep{harper2005evolution,hewitt2013seasonal}, which is simplified to a static water pressure in the current model.

The second main limitation is that the model does not account for the three-dimensional effects of sticky patches. In a real ice sheet, some sticky patches may have a round shape instead of a long, rib-like pattern. In that case, it is more appropriate to study them in a 3-D domain, which includes both vertical and lateral extension of cracks, rather than as a plane-strain problem in the $x$--$z$ plane.

The third limitation relates to the fracture toughness. In the LEFM approach, in order to determine the maximum crack length, we assume a constant fracture toughness $K_{C}$, neglecting any spatial variation due to its dependence on grain size \citep{schulson1984brittle}. However, regions and layers of coarse-grained ice exist in ice sheets \citep{gow1997physical}.
Thus, fracture toughness in some glaciers and ice sheets may be a function of depth or lateral position. This heterogeneity would modify the predicted crack lengths. A more complete understanding of how basal crevasses grow and interact with viscous flow will require a three-dimensional, viscoelastic model including variations in fracture toughness.

\section{Conclusions}
Besides basal water pressure \citep{van1998fracture}, stress variations of sticky patches at the ice--bed interface can promote the initiation of basal crevasses. Prior to this study, basal crevasses were assumed to be mode-I hydrofractures under pure horizontal tension \citep{van1998fracture,krug2014combining,jimenez2018evaluation}. Assuming water pressure smaller than the flotation condition, we examined the effect of sticky patches on basal crevassing in a grounded glacier or ice sheet. We found that sticky patches can provide stress required for crack extension. Alongside the flotation fraction, such sticky-patch-assisted crevassing depends on two non-dimensional parameters: (1) the ratio of the sticky-patch half-width to the ice thickness, $W^{\prime}$, and (2) the ratio of excess shear stress to the basal lithostaic pressure, $\Delta\tau^{\prime}$.

With a sufficient variation of basal shear stress, the direction of basal fracture is controlled by the relative size of the sticky patch. When the width of the sticky patch is much larger than the ice thickness, basal crevasses grow nearly vertically and are essentially mode-I fractures. When the width of the sticky patch is smaller than the ice thickness, however, curved basal crevasses grow with trajectories inclined upstream. In this case, principal stresses can be used to approximate crack trajectories. 

For real glaciers or ice sheets with complicated geometries, our model can be combined with the basal stress pattern to investigate how stress variation promotes basal crevassing. Compared to stress estimates from inversion calculations \citep{sergienko2013regular,braedstrup2016basal}, qualitative analysis shows that for a flotation fraction of about $0.9$, the shear-stress pattern in alpine glaciers and ice sheets may play an important role in determining the growth of basal crevasses. To better understand the basal crevasses controlled by sticky patches, future research could incorporate viscous ice flow, spatially resolved subglacial hydrology, and detailed fracturing properties of ice.

\appendix
\section{Appendix A. 2D elasticity in terms of perturbation stress}\label{sec:AppendixA}
\setcounter{equation}{0}
\renewcommand\theequation{A\arabic{equation}}
 The domain is a notched ice strip with length $2L$ as shown in Figure~\ref{fig:Schematic}. We set $L\gg W$ to make sure the crack is far from the boundaries to avoid any edge effects. Substituting $\boldsymbol{T}=\boldsymbol{\sigma}+p_{i}\boldsymbol{I}$ into the governing equation and boundary conditions, we can get stress equilibrium equation in terms of the perturbation stress as
\begin{equation}\label{eq:governing_p}
    \boldsymbol{\nabla}\cdot\boldsymbol{T}=\boldsymbol 0.
\end{equation}
For the top boundary, a traction-free boundary condition is imposed as
\begin{equation}\label{eq:boundary_top_p}
    \boldsymbol{T}\cdot\boldsymbol{n}=\boldsymbol{0} \quad \text{for}\quad z=H,
\end{equation}
where $\boldsymbol{n}$ is the outward-pointing unit normal vector of the domain. On both sides, we impose traction-free boundary condition for the perturbation stress,
\begin{equation}\label{eq:boundary_sides_p}
    \boldsymbol{T}\cdot\boldsymbol{n}=\boldsymbol{0} \quad\text{for}\quad x=\pm L.
\end{equation}
On the crack walls ($x=W$, $0\le z\le Z_{C}$), compression caused by static water pressure is imposed as
\begin{equation}\label{eq:boundary_crack_p}
    \boldsymbol{T}\cdot\boldsymbol{n} = \left\{
    \begin{aligned}
    &\rho_{i}g H\left[\left(\frac{\rho_{w}}{\rho_{i}}-1\right)\frac{z}{H} + 1 - f\right]\boldsymbol{n}\quad\text{for}\quad z < H_w,\\
    & \rho_{i}g H\left(-\frac{z}{H} + 1\right)\boldsymbol{n}\quad\text{for}\quad z\ge H_w,
    \end{aligned}
    \right.
\end{equation}
where ${\rho_i}fH/{\rho_w}=H_w$ is the hydraulic head. The bottom boundary condition remains the same as \eqref{eq:bc_bottom_x} and \eqref{eq:bc_bottom_z}, since the overburden stress doesn't contribute to the shear stress and the elastic deformation,
\begin{align}\label{eq:boundary_bottom1_p}
    \boldsymbol{t}\cdot\boldsymbol{T}\cdot\boldsymbol{n} &= \tau\left(x,0\right)=\left\{
    \begin{aligned}
    &-\frac{W}{L}\Delta\tau\quad&\left|x\right|\geq W, \\
    &\left(1-\frac{W}{L}\right)\Delta\tau\quad &\left|x\right|< W,
    \end{aligned}
    \right.\\
    \boldsymbol{u}\cdot\boldsymbol{n} &= 0.
\end{align}
Here $\boldsymbol{t}=-\bvec{x}$ is the unit tangent vector to the bottom boundary. Note that in equation \eqref{eq:boundary_bottom1_p} an extra term $-W\Delta\tau/L$ is added on the bottom boundary in order to maintain the total force balance of the ice strip. Because $L\gg W$, we can neglect the near-field effect of this force-balance term.

Different from standard elasticity, we set up constitutive relation between the $\textit{perturbation}$ stress tensor $\boldsymbol{T}$ and the strain tensor $\boldsymbol{\epsilon}$ as in equation~\eqref{eq:constitutive}.

\section{Appendix B. Benchmark}\label{sec:AppendixB}
\setcounter{equation}{0}
\renewcommand\theequation{C\arabic{equation}}
The weight function method has proved to be a useful tool to calculate stress intensity factors in ice sheets under certain basal boundary conditions \citep{tada2000analysis, jimenez2018evaluation}. The essence is calculating stress intensity factors by integrating the stress along a hypothesised crack in an uncracked domain with an appropriate weight function. The advantage of the weight function method is that the computation is performed in the uncracked domain with no discontinuity and singularity caused by cracks. For certain simple domain and crack geometry and boundary conditions, the weight function method can give SIFs with high accuracy.

\cite{jimenez2018evaluation} has provided the appropriate weight function from \cite{tada2000analysis} to be used for basal cracks in a grounded ice-sheet:
\begin{equation}\label{eq:modeI_integral}
    K_{\RNum{1}}=\int_{0}^{Z_{C}}\sigma_{xx} \left( z \right)G_1\left(\lambda,\gamma\right)\text{d}z,
\end{equation}

\begin{equation}\label{eq:modeI_weight}
    G_1\left(\lambda,\gamma\right)=\frac{2}{\sqrt{2H}}\sqrt{\frac{\tan\left(\frac{\pi\lambda}{2}\right)}{1-{\cos\frac{\pi\lambda}{2}}/{\cos\frac{\pi\lambda\gamma}{2}}}}
    \left\{1+0.297\sqrt{1-\gamma^2}\left[1-\cos\left(\frac{\pi}{2}\lambda\right)\right]\right\},
\end{equation}
where $\lambda=Z_{C}/{H}$ is the non-dimensional crack length and $\gamma=z/Z_C$ is the $z$ coordinate normalised by the crack length. This weight function can be used to predict the mode-I stress intensity factor $K_{\RNum{1}}$ in grounded ice. A similar weight function in \cite{tada2000analysis} can also be applied to calculating mode-II stress intensity factor:
\begin{equation}\label{eq:modeII_integral}
    K_{\RNum{2}}=\int_{0}^{Z_{C}}\sigma_{xz} \left( z \right)G_2\left(\lambda,\gamma\right)\text{d}z,
\end{equation}

\begin{equation}\label{eq:modeII_weight}
    G_2\left(\lambda,\gamma\right)=\frac{2}{\sqrt{2H}}\sqrt{\frac{\tan\left(\frac{\pi\lambda}{2}\right)}{1-{\cos\frac{\pi\lambda}{2}}/{\cos\frac{\pi\lambda\gamma}{2}}}}
    \left\{1+0.297\sqrt{1-\gamma^2}\left[1-\cos\left(\frac{\pi}{2}\lambda\right)\right]\right\}\frac{\sin{\frac{\pi \lambda\gamma}{2}}}{\sin{\frac{\pi \lambda}{2}}}.
\end{equation}

\begin{figure}
    \centering{\includegraphics[width=0.4\textwidth]{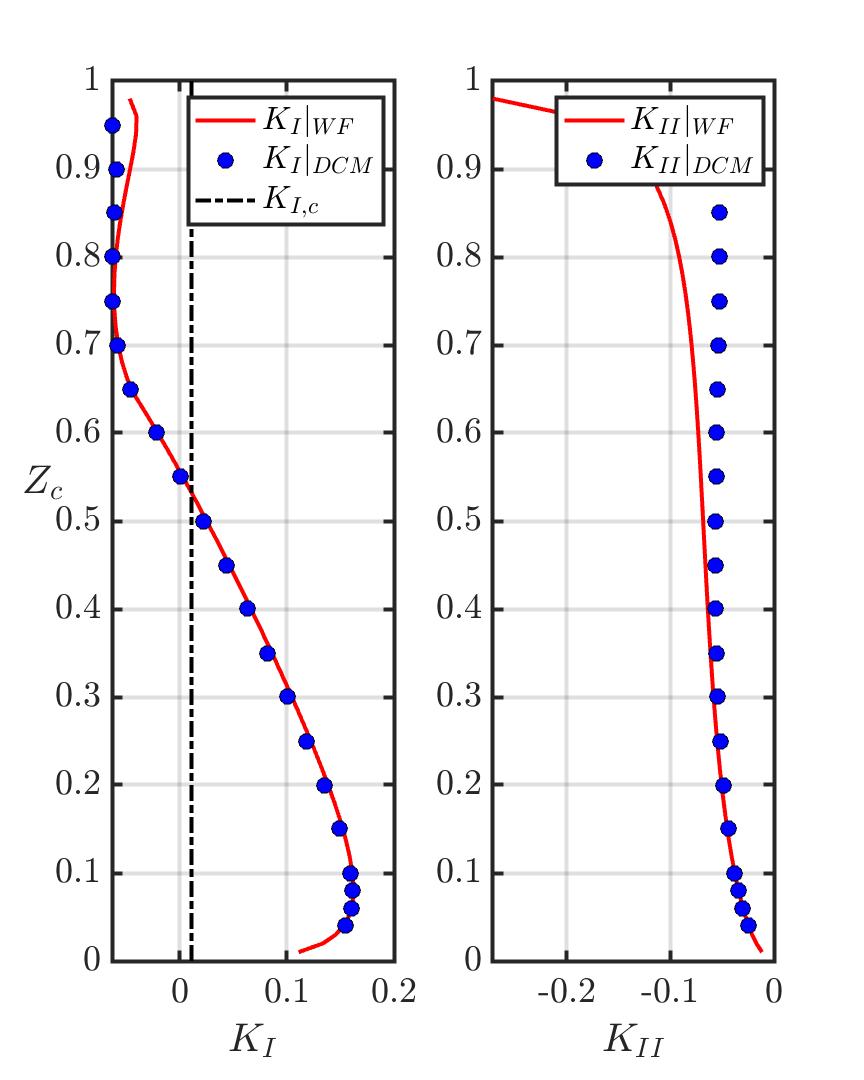}}
    \caption{Stress intensity factors $K_{\RNum{1}}$ and $K_{\RNum{2}}$ when $\Delta\tau=0.3$ and $f=0.7$. The red lines show SIFs calculated by the weight function method. The blue dots show SIFs calculated by DCM.}
    \label{fig:K1_WF_DCM}
\end{figure}

To verify the implementation of the DCM method we consider a problem where there is a mixed-mode (mode-I and mode-II) water-filled crack. SIFs are calculated by both the DCM method and the weight function method in Figure~\ref{fig:K1_WF_DCM}. In the DCM calculation, the element size is set to be $0.005$ times the crack length near the crack tip and $0.2$ ice thicknesses away from the tip.

For $K_{\RNum{1}}$, these two methods agree over most crack lengths. However, there is large deviation for $K_{\RNum{2}}$ at large crack length.

\section{Appendix C. Thermal Structure of basal crevasses}\label{sec:AppendixC}
\setcounter{equation}{0}
\renewcommand\theequation{E\arabic{equation}}
\subsection{Thermal structure of a single basal crevasse}
Refreezing of water inside basal crevasses is a potential factor affecting the thermal profile of ice. Using some analytical solutions of \cite{carslaw1959conduction}, this section shows how a single basal crevasse or a series of basal crevasses affect the thermal structure of ice.

In order to simplify the computation, we neglect the advection and other englacial heat sources and focus on heat diffusion after refreezing \citep{luckman2012basal}. The background temperature profile is assumed to have no effect on the heat conduction caused by refreezing. That means that the simple analytical model just accounts for the effect of refreezing and doesn't make any prediction of the net temperature profile. Let $\Delta T$ be the temperature perturbation induced by basal crevasse. The ice sheet is simplified to an isotropic infinite strip $0<z<H$. For boundaries, we assume zero temperature perturbation ($\Delta T=0$) at $z=0$, $z=H$ and $x=\pm \infty$. The governing equation of thermal conduction in terms of $\Delta T$ is
\begin{equation}
    \frac{\partial^2 \Delta T}{\partial x^2}+\frac{\partial^2 \Delta T}{\partial z^2} = \frac{1}{\kappa}\frac{\partial \Delta T}{\partial t}.
\end{equation}
The boundary conditions are
\begin{gather}
    \Delta T\left(\pm \infty,z\right)=0,\\
    \Delta T\left(x,0\right)=\Delta T\left(x,H\right)=0.
\end{gather}
Table \ref{constants} shows the values of parameters used in the calculation. At time $t=0$, an amount of heat $q_{i}$ per unit length per unit depth into the page is released by a line heat source from $\left(0,0\right)$ to $\left(0,Z_{c}\right)$. In order to get an estimated value of that released heat, we need to estimate the volume of water refreezes at $t=0$. Based on the BEM simulations, the width of the crack $w$ is approximately $0.1$~m, $q_{i}$ is estimated as
\begin{equation}\label{eq:q_i}
    q_{i}=\rho_{i} L w= 3\times 10^{7}\ \text{J}\cdot \text{m}^{-2},
\end{equation}
where $L$ is the latent heat of fusion of water to ice. 

Thus we have the full thermal problem in an infinite strip with an initial line heat source representing refreezing in real basal crevasses.
\begin{table}
\centering
\caption{Constants used in calculation of temperature around basal crevasses.}
\label{constants}

\begin{tabular}{@{}lcc}\hline
Physical property
  & Notation
  & Value \\ \hline
Density of water & $\rho_w$ & $1.0\times10^3$ $\text{kg}\cdot \text{m}^{-3}$\\
Density of ice & $\rho_i$ & $0.92\times10^3$ $\text{kg}\cdot \text{m}^{-3}$\\
Latent heat of fusion & $L$ & $3.34\times 10^5$ $\text{J}\cdot \text{kg}^{-1}$\\
Thermal Conductivity & $k$ & $2.1$ $\text{W}\cdot \text{m}^{-1}\text{K}^{-1}$\\
Heat capacity of ice & $C_{p}$ & $2.10\times 10^{3}$ $\text{J}\cdot \text{kg}^{-1}\cdot \text{K}^{-1}$\\
Thermal diffusivity & $\kappa$ & $1.09\times 10^{-6}$ $\text{m}^2\cdot \text{s}^{-1}$\\
Sliding velocity & $v_{s}$ & $1.0\times10^2$ $\text{m}\cdot \text{y}^{-1}$\\
Crevasse spacing & $w_{s}$ & $1.0\times10^2$ $\text{m}$

\end{tabular}
\end{table}
The solutions are given by \cite{carslaw1959conduction} as
\begin{equation}\label{eq:greens_tem}
    \Delta T = \frac{q_i}{\pi\rho_i C_p\sqrt{\kappa \pi t}}\exp{\left(-\frac{\left(x-x^{\prime}\right)^2}{4\kappa t}\right)} \sum_{{n=1}}^{+\infty}\frac{1}{n}\left(1-\cos\left({\frac{n\pi Z_{C}}{H}}\right)\right)\sin\left({\frac{n\pi z}{H}}\right)\exp\left({\frac{-\kappa n^2 \pi^2 t}{H^2}}\right).
\end{equation}
\subsection{Thermal structure of a series of equally spaced basal crevasses}
A stable sticky patch can cause a local stress variation and potentially generate a series of basal crevasses. The thermal effect of a single basal crevasse, which is shown above, can be advected downstream and superposed with the effect of other basal crevasses, resulting in temperature variation on a longer timescale and larger area. Mathematically, based on $\Delta T\left(x,z,t\right)$ that we have obtained above, the net effect of a series of basal crevasses can be obtained by a simple linear superposition of the $\Delta T$ of each crevasse.
\begin{figure}
    \centering{\includegraphics[width=0.5\textwidth]{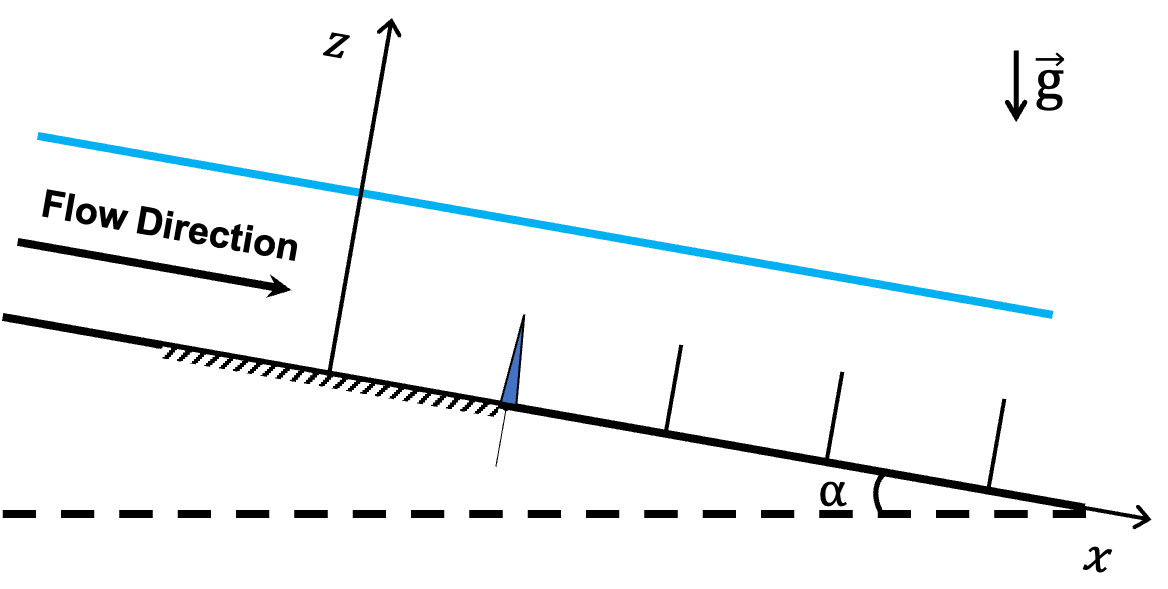}}
    \caption{Schematic of a series of basal crevasses produced on a sticky patch.}
    \label{fig:heat_series}
\end{figure}
Assuming these crevasses are generated at a fixed spacing $w_{s}=W=100\ \text{m}$ in an $100$-m-thick ice sheet, starting from $t=0$, we fix the coordinate system on the sticky patch and move the ice sheet at $v_{s}=100$ $\text{m}\times\text{y}^{-1}$. Thus the crevasses are also generated at a fixed time interval $\Delta t=W/v_s=1$ y. The net temperature perturbation $\Delta T_{net}$ is
\begin{equation}
    \Delta T_{net}=\sum_{n=0}^{\infty} \Delta  T\left(x-x_{n}^{\prime},\ z,\ t-t_{n}^{\prime}\right),
\end{equation}
where $n$ is the number of the crack, $x_{n}^{\prime}=n w_{s}$ is the position of the crack $n$, $t_{n}^{\prime}=-n\Delta t$ is the time when the crack $n$ formed and refroze.
\section{Code availability}
We use the stable version of FEniCS \citep{logg2010dolfin,logg2012automated,LangtangenLogg2017} (\href{https://quay.io/repository/fenicsproject/stable}{https://quay.io/repository/fenicsproject/stable}) and the open-source boundary element code CutAndDisplace (\href{https://github.com/Timmmdavis/CutAndDisplace.git}{https://github.com/Timmmdavis/CutAndDisplace.git}) developed by \cite{davis2017new}. 
The code for the runs in this study is available at the online repository \linebreak(\href{https://doi.org/10.5281/zenodo.6364838}{https://doi.org/10.5281/zenodo.6364838}).

\section{Acknowledgements}

We thank B.~Lipovsky and C.-Y.~Lai for their guidance on the application of LEFM to ice. The general idea that we explore arose during a conversation with P.~Christoffersen and R.~Law. This research received funding from the European Research Council under Horizon 2020 research and innovation program grant agreement number 772255.

\bibliography{igsrefs}   
\bibliographystyle{igs}  

\end{document}